\documentclass[12pt,a4paper]{conference}

\usepackage{fancyhdr}
\usepackage{graphicx,amsmath,amssymb,cite}
\usepackage{multind}

\newcommand{\beq}{\begin{equation}}
\newcommand{\eeq}[1]{\label{#1}\end{equation}}
\newcommand{\eeqn}{\end{equation}}

\newcommand{\beqa}{\begin{eqnarray}}
\newcommand{\eeqa}[1]{\label{#1}\end{eqnarray}}
\newcommand{\eeqan}{\end{eqnarray}}

\let\bar=\overbar

\newcommand{\Dslash}{\not{\hbox{\kern-4pt $D$}}}
\newcommand{\dslash}{\not{\hbox{\kern-2pt $\del$}}}

\newcommand{\msb}{{\bar{\ssstyle M \kern -1pt S}}}

\begin{document}

\Chapter{RENORMALIZING THE SCHR\"ODINGER EQUATION FOR NN SCATTERING}
           {Renormalization of NN scattering}{E. Ruiz Arriola \it{et al.}}
\vspace{-6 cm}\includegraphics[width=6 cm]{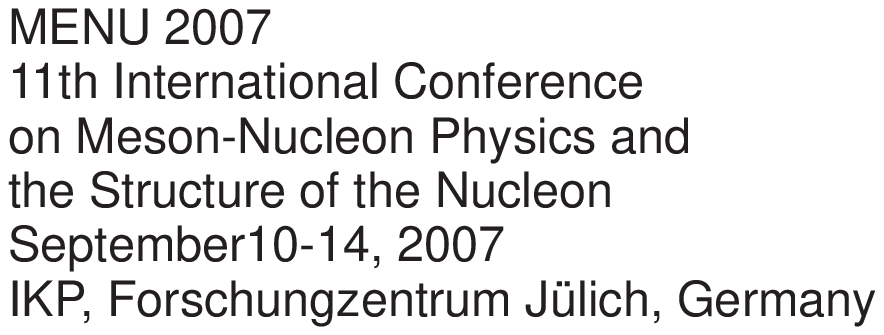}
\vspace{4 cm}

\addcontentsline{toc}{chapter}{{\it N. Author}} \label{authorStart}

\begin{raggedright}

{\it \underline{E. Ruiz Arriola}~\footnote{Invited Speaker at Menu
2007.}, A. Calle Cord\'on, M. Pav\'on Valderrama\footnote{Institut f\"ur Kernphysik, Forschungszentrum
J\"ulich, 52425 J\"ulich, Germany}}\index{author}{Author, N.}\\
Departamento de F\'{\i}sica At\'omica, Molecular y Nuclear \\
Universidad de Granada \\ E-18071 Granada, Spain.

\bigskip\bigskip

\end{raggedright}

\begin{center}
\textbf{Abstract}
\end{center}
The renormalization of the Schr\"odinger equation with regular One
Boson Exchange and singular chiral potentials including One and
Two-Pion exchanges is analyzed within the context of NN scattering.

\section{Introduction}

One traditional view of NN force has been through One Boson Exchange
(OBE) Models~\cite{Machleidt:1987hj,Machleidt:2000ge}. Recent
developments have shown how chiral symmetry may provide NN forces of
practical interest in nuclear
physics~\cite{Bedaque:2002mn,Machleidt:2005uz,Epelbaum:2005pn}. Remarkably,
chiral expansions, based on assuming a large scale suppression on the
parameters $ 4 \pi f_\pi \sim M_N \sim 1 {\rm GeV}$ necessarily
involve singular potentials at short distances, i.e. $r^2 |V(r)| \to
\infty $ for $r\to 0$. If we take the limit $r \ll 1/m_\pi $ (or
equivalently large momenta) pion mass effects are irrelevant and hence
at some fixed order of the expansion one has
\begin{eqnarray}
 V(r)  \sim  \frac{M_N}{(4 \pi f_\pi)^{2 n}  M_N^m} \frac{1}{r^{2 n+m}} 
\label{eq:pot-sing}
\end{eqnarray}
(the only exception is the singlet channel-OPE case which behaves as
$\sim m_\pi^2/ f_\pi^2 r$, see below). The dimensional argument is
reproduced by loop calculations in the so called Weinberg dimensional
power counting~\cite{Kaiser:1997mw,Rentmeester:1999vw}.  Thus, much of
our understanding on the physics deduced from chiral potentials might
be related to a proper interpretation of these highly singular
potentials. Renormalization is the most natural tool provided 1) we
expect the potential is realistic at long distances and 2) we want
short distance details not to be essential in the description. This is
precisely the situation we face most often in nuclear
physics. Knowledge on the attractive or repulsive character of the
singularity turns out to be crucial to successfully achieve this
program and ultimately depends on the particular scheme or power
counting used to compute the potential. We illustrate our points for
the simpler OBE potential in the $^1S_0$ channel and then review some
results for chiral OPE and TPE potentials for all partial waves and
the deuteron bound state.

\section{Renormalization of OBE potentials}

The singularity of chiral potentials raises suspicions and, quite
often, much confusion. However, if properly interpreted and handled
they do not differ much from the standard well-behaved regular
potentials one usually encounters in nuclear physics.  Actually, we
digress here that renormalization may provide useful insights even if
the potential is not singular at the origin ($r^2 V(r) \to 0$ !). For
definiteness, let us analyze as an illustrative example the phenomenologically
successful $^1S_0$ OBE
potential~\cite{Machleidt:2000ge,Machleidt:1987hj} (we take $m_\rho =
m_\omega$)
\begin{eqnarray}
V(r) = - \frac{g_{\pi NN}^2 m_\pi^2}{16 \pi M_N^2} \frac{e^{-m_\pi
r}}{r } -\frac{g_{\sigma NN}^2}{4\pi}
\frac{e^{-m_\sigma r}}{r} +\frac{g_{\omega NN}^2}{4\pi}
\frac{e^{-m_\omega r}}{r} + \dots 
\label{eq:pot}
\end{eqnarray} 
where for simplicity we neglect nucleon mass effects and a tiny $\eta$
contribution. We take $m_\pi=138 {\rm MeV}$, $M_N=939 {\rm MeV}$,
$m_\omega=783 {\rm MeV}$ and $g_{\pi NN}=13.1$ which seem firmly
established. Actually, Eq.~(\ref{eq:pot}) looks like a long distance
expansion of the potential. NN scattering in the elastic region below
pion production threshold involves CM momenta $p < p_{\rm max} = 400
{\rm MeV}$. Given the fact that $1/m_\omega=0.25 {\rm fm} \ll 1/p_{\rm
max}=0.5 {\rm fm}$ we expect heavier mesons to be irrelevant, and
$\omega$ itself to be marginally important. In the traditional
approach, however, this is not
so~\cite{Machleidt:2000ge,Machleidt:1987hj}.  Actually, the problem is
essentially handled by solving the Schr\"odinger equation (S-wave)
\begin{eqnarray} 
-u_p''(r) + M V(r) u_p(r) = p^2 u_p(r) 
\label{eq:up}
\end{eqnarray} 
with the {\it regular} solution at the origin, $u_p(0)=0$. This
boundary condition implicitly assumes taking also the potential all
the way down to the origin.  The asymptotic condition for $r \gg
1/m_\pi$ is taken to be
\begin{eqnarray} 
u_p (r) \to \frac{\sin ( p r + \delta_0 (p))}{\sin \delta_0 (p)}
\label{eq:d0}
\end{eqnarray} 
where $\delta_0(p)$ is the phase-shift. For the potential in
Eq.~(\ref{eq:pot}) the phase shift is an analytic function of $p$ with
the closest branch cut located at $p = \pm i m_\pi /2 $, so that
one can undertake an effective range expansion,
\begin{eqnarray}
p \cot \delta_0 (p) = - \frac1{\alpha_0} + \frac12 r_0 p^2 + v_2 p^4 + \dots 
\label{eq:ere}
\end{eqnarray} 
within a radius of convergence $|p| \le m_\pi /2$. 
A similar expansion for the wave function $u_p (r)= u_0 (r) + p^2 u_2 (r) \dots $ means solving the set of equations 
\begin{eqnarray} 
-u_0''(r) + M V(r) u_0(r) &=& 0 \, , \label{eq:u0} \\  
u_0 (r) &\to& 1- r /\alpha_0  \, ,  \nonumber \\ 
-u_2 '' (r) + U(r) u_2 (r) &=& u_0 (r) \, , \label{eq:u2} \\
u_2 (r) &\to& \left(r^3
-3 \alpha_0 r^2 + 3 \alpha_0 r_0 r \right)/(6 \alpha_0) \, , \nonumber
\end{eqnarray} 
where, again, the {\it regular} solutions, $u_0(0)=u_2(0)= 0$
are taken.  With this normalization the effective range $r_0$ is computed
from the standard formula
\begin{eqnarray}
r_0 = 2\int_0^\infty dr \left[  \left(1- r
/\alpha_0\right)^2 - u_0 (r)^2\right] \, .
\label{eq:r0}
\end{eqnarray} 
In the usual approach~\cite{Machleidt:1987hj,Machleidt:2000ge} {\it
everything} is obtained from the potential assumed to be valid for $0
\le r < \infty$. In practice, strong form factors are included
mimicking the finite nucleon size and reducing the short distance
repulsion of the potential, but the regular boundary condition is
always kept.~\footnote{Calculations solving the equivalent
Lippmann-Schwinger equation in momentum space for regular potentials
correspond always to choose the regular solution for the Schr\"odinger
equation in coordinate space.}  As it is well known the $^1S_0$
scattering length is unnaturally large $\alpha_0 = -23.74(2) {\rm
fm}$, while $r_0= 2.77(4) {\rm fm}$. Let us assume we have fitted the
potential, Eq.~(\ref{eq:pot}), to reproduce $\alpha_0$. Under these
circumstances a tiny change in the potential $V \to V + \Delta V$ has
a dramatic effect on $\alpha_0$, since one obtains
\begin{eqnarray}
\Delta \alpha_0 = \alpha_0^2 M_N \int_0^\infty \Delta V(r) u_0(r)^2 dr  \, .
\label{eq:delta-alpha0} 
\end{eqnarray} 
As a result, potential parameters must be fine tuned. In particular,
the resulting $\omega$-repulsive contribution is well determined with
an unnaturally large coupling, $g_{\omega NN} \sim
16$.~\cite{Machleidt:1987hj,Machleidt:2000ge}. In our case, with no
form factors nor relativistic corrections, a fit to
Ref.~\cite{Stoks:1994wp} yields $g_{\omega NN} = 12.876(2)$,
$g_{\sigma NN} = 12.965(2)$ and $m_\sigma = 554.0(4) {\rm MeV}$ with
$\chi^2 /{\rm DOF}=0.26$. Note the small uncertainties, as expected
from our discussion and Eq.~(\ref{eq:delta-alpha0}).  As mentioned
above $1/m_\omega=0.25 {\rm fm} \ll 1/p_{\rm max}=0.5 {\rm fm}$ so
$\omega$ should not be crucial at least for CM momenta $p \ll p_{\rm
max}$. Thus, despite the undeniable success in fitting the data this
sensitivity to short distances looks counterintuitive.

\begin{figure}[tbc]
\begin{center}
\includegraphics[height=6.5 cm,width=4.5 cm,angle=270]{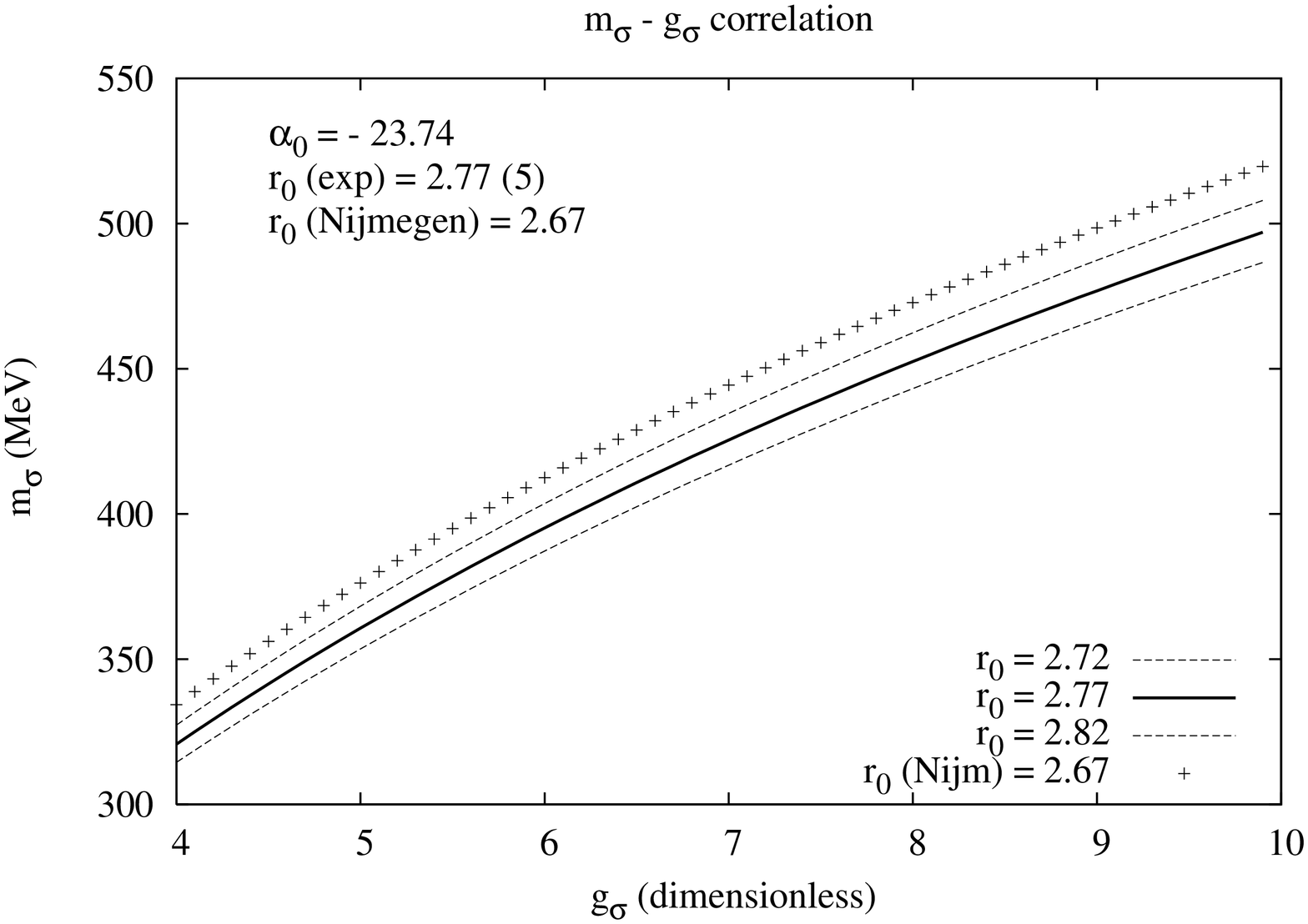}
\includegraphics[height=6.5 cm,width=4.5
cm,angle=270]{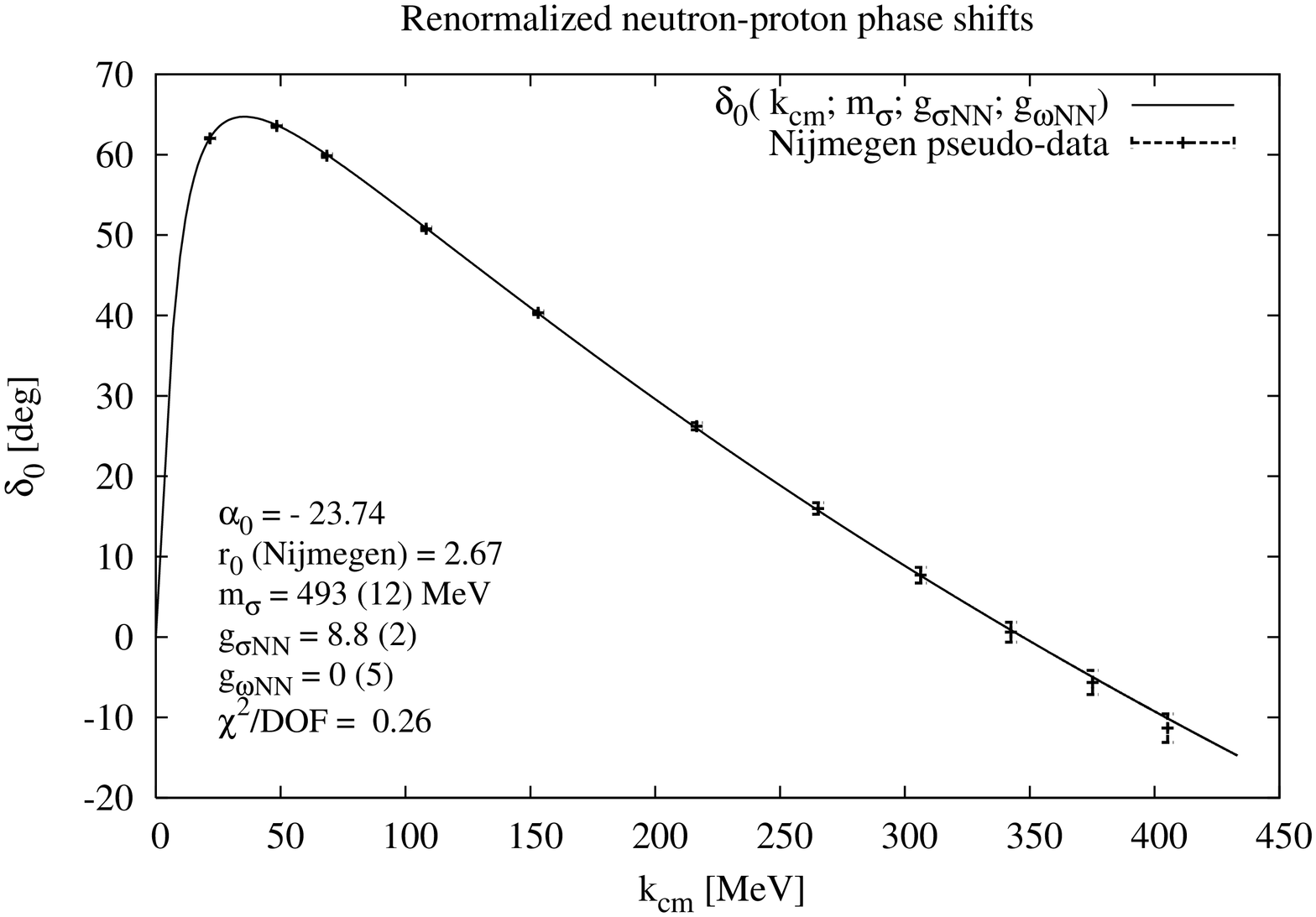}\caption{Results in the $^1S_0$ channel
for the renormalized OBE potential. Left: Effective range correlation
between $g_{\sigma NN}$ and $m_\sigma$ for $g_{\omega NN}=0$. Right:
Renormalized phase shift (in degrees) as a function of the CM momentum
(in MeV) in the OBE ($\pi+\sigma+\omega$) model.  The data are an
average of \cite{Stoks:1994wp}.}
\label{fig:sigma}
\end{center}
\end{figure}

The renormalization viewpoint {\it refuses} to access physically the
very short distance region, but encodes it through low energy
parameters described by the effective range expansion,
Eq.~(\ref{eq:ere}), as renormalization conditions (RC's). In the case
of only one RC where $\alpha_0$ is fixed one proceeds as follows~\cite{PavonValderrama:2003np,OPE+TPE}:
\begin{itemize}  

\item For a given $\alpha_0$ integrate in the zero energy wave
function $u_0(r)$, Eq.~(\ref{eq:u0}), down to the cut-off radius
$r_c$. This is the RC.

\item Implement self-adjointness through the boundary
condition
\begin{eqnarray} 
u_p'(r_c)  u_0 (r_c) - u_0'(r_c) u_p(r_c) =0 \, , 
\label{eq:boundary}
\end{eqnarray} 

\item Integrate out the finite energy wave function $u_p(r)$, from Eq.~(\ref{eq:up}) and  determine the phase shift $\delta_0(p)$ from
Eq.(\ref{eq:d0}). 

\item Remove the cut-off $r_c \to 0$ to strive for model independence.
\end{itemize} 
This allows to compute $\delta_0$ (and hence $r_0$, $v_2$ ) from
$V(r)$ and $\alpha_0$ as {\it independent} information. Note that this
is equivalent to consider, in addition to the regular solution, the
{\it irregular} one~\footnote{In momentum space this can be shown to
be equivalent to introduce one counterterm in the cut-off
Lippmann-Schwinger equation, see Ref.~\cite{Entem:2007jg} for a
detailed discussion.}. A beautiful result is the universal low energy
theorem which highlights this de-correlation between the potential and
the scattering length~\cite{OPE+TPE}
\begin{eqnarray}
r_0 = 2 \int_0^\infty dr ( 1 - u_{0,0}^2 ) -\frac{4}{\alpha_0}
\int_0^\infty dr ( r - u_{0,0} u_{0,1} ) + \frac{2}{\alpha_0^2} \int_0^\infty
dr ( r^2 - u_{0,1}^2 ) ,
\end{eqnarray} 
based on the superposition principle of boundary conditions, i.e.
writing $u_0 (r) = u_{0,0} (r) - u_{0,1} (r)/\alpha_0$ with
$u_{0,n}(r) \to r^n $ and using Eq.(\ref{eq:r0}). A fit of the
potential (\ref{eq:pot}) with $g_{\omega NN}=0$ to the effective range
yields (Fig.~\ref{fig:sigma}) a strong correlation between $m_\sigma$
and $g_{\sigma NN}$. Over-imposing this correlation to $r_0 = 2.670(4)
{\rm fm}$, a fit to Ref.~\cite{Stoks:1994wp} yields $ m_\sigma =
493(12) {\rm MeV} $, $g_{\sigma NN}= 8.8(2) $, $g_{\omega NN}= 0(5) $
with $\chi^2 /{\rm DOF} = 0.24 $ (Fig.~\ref{fig:sigma}). Note the
larger uncertainties, although correlations allow $g_{\omega NN} \sim
9 $ and $m_\sigma \sim 520{\rm MeV}$ within $\Delta \chi^2
=1$. Contrary to common wisdom, but according to our naive
expectations, no strong short range repulsion is essential. The moral
is that building $\alpha_0$ {\it from} the potential is
equivalent to absolute knowledge at short distances and in the $^1S_0$
channel a strong fine tuning is at work. Of course, a more systematic
analysis should be pursued in all partial waves and relativistic
corrections might be included as well, but this example illustrates
our point that the renormalization viewpoint may tell us to what
extent short distance physics may be less well determined than the
traditional approach assumes. This opens up a new perspective to
the phenomenology of OBE potentials in cases where the strong
$\omega$-repulsion has proven to be crucial at low
energies~\cite{ACC07}.

\section{Renormalization of chiral potentials}

\begin{figure}[ttt]
\begin{center}
\includegraphics[height=4.5 cm,width=6.5 cm]{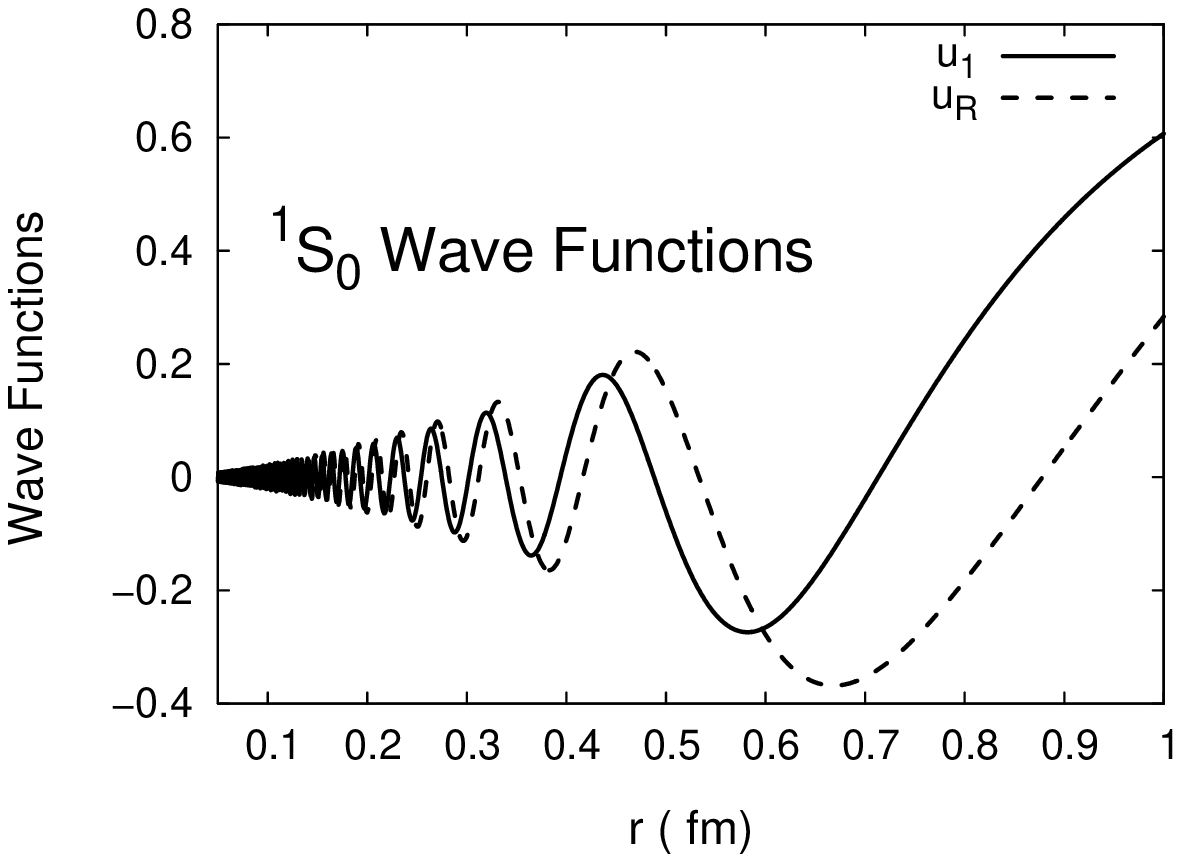} 
\includegraphics[height=4.5 cm,width=6.5 cm]{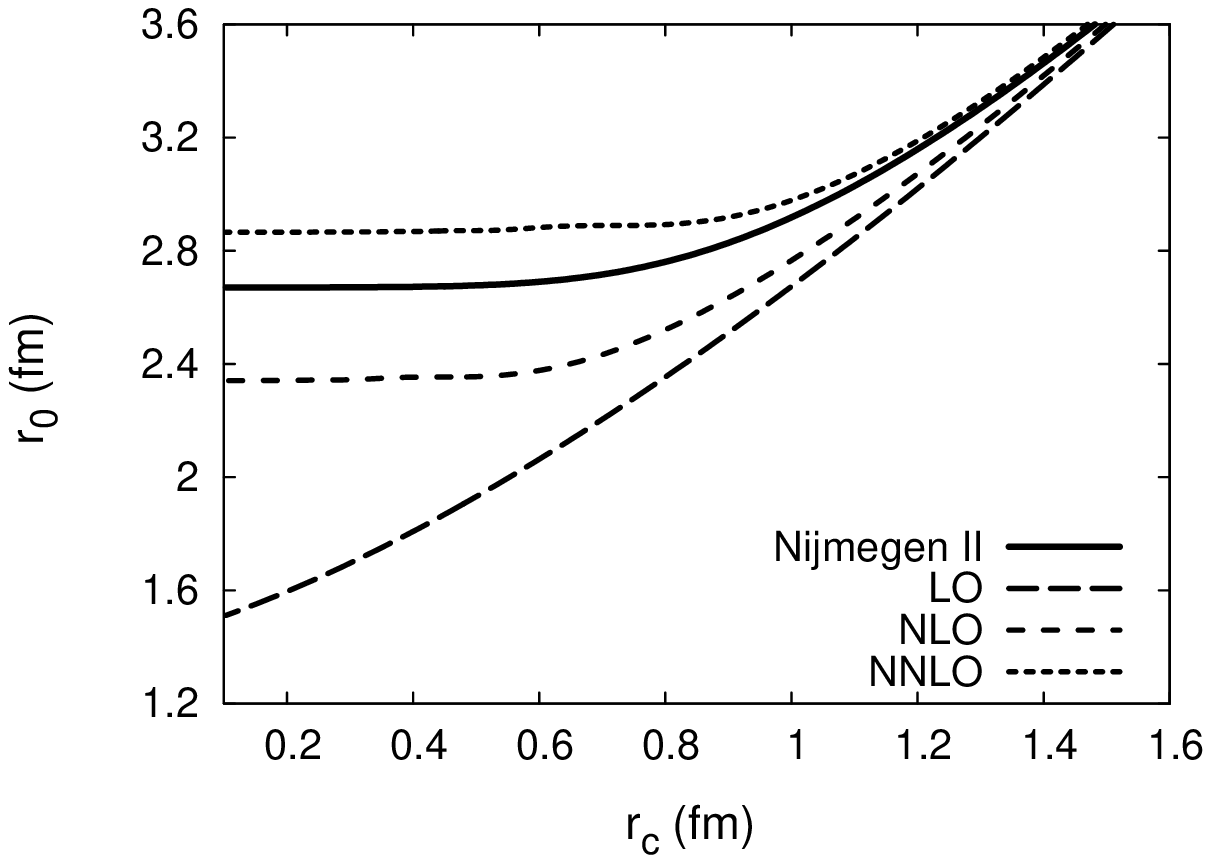}
\caption{ (Left panel) Zero-energy, $^1S_0$ linearly independent wave
functions at NNLO; $u_1 \to 1 $ and $u_r \to r $ for $r \to \infty$.
(Right Panel) Effective range $r_0$ as a function to the cut-off for
the same channel and different orders; using $r_0(r_c) =
2\,\left(\int_0^{\infty}(1-r/\alpha_0)^2 \,dr
-\int_{r_c}^{\infty}u_0^2\,dr \right) $, with $\alpha_0 = -23.74 {\rm
fm}$~\cite{Stoks:1994wp}.}
\label{fig:s-wave}
\end{center}
\end{figure}

\begin{figure}[tbc]
\begin{center}
\includegraphics[height=5 cm,width=6.5 cm]{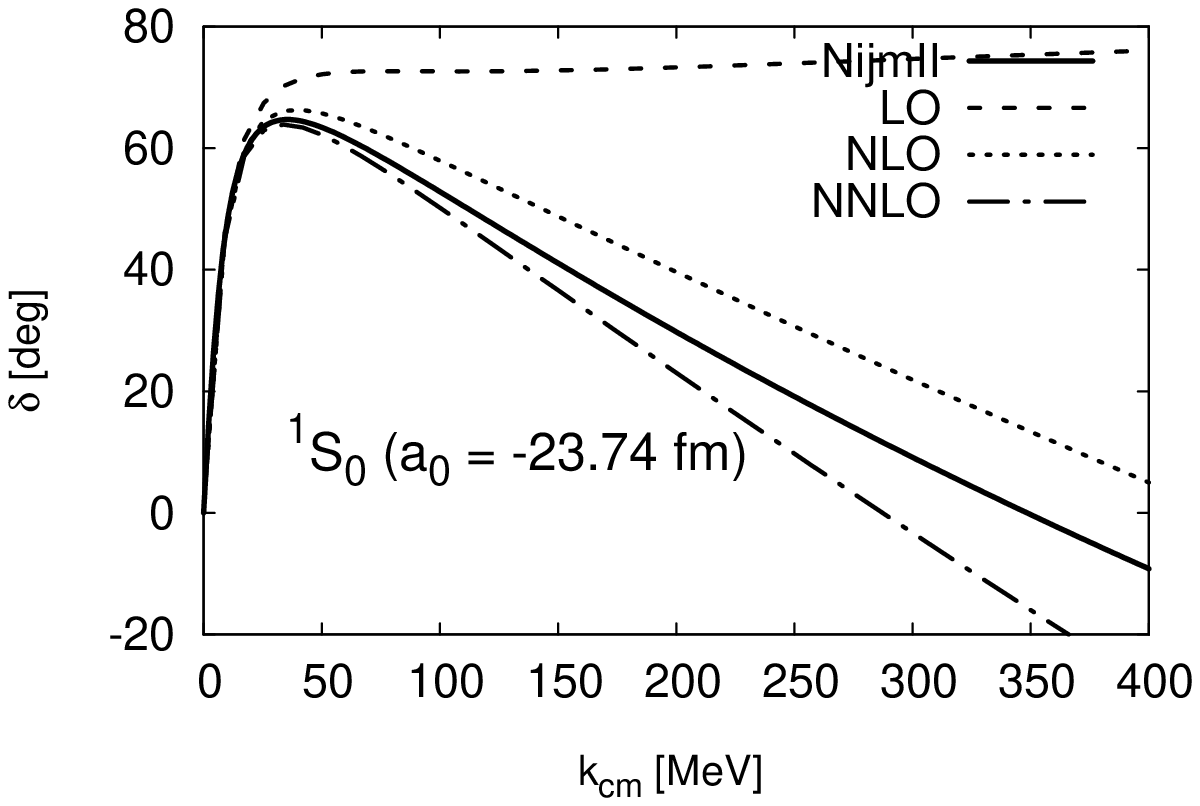}
\includegraphics[height=5 cm,width=6.5 cm]{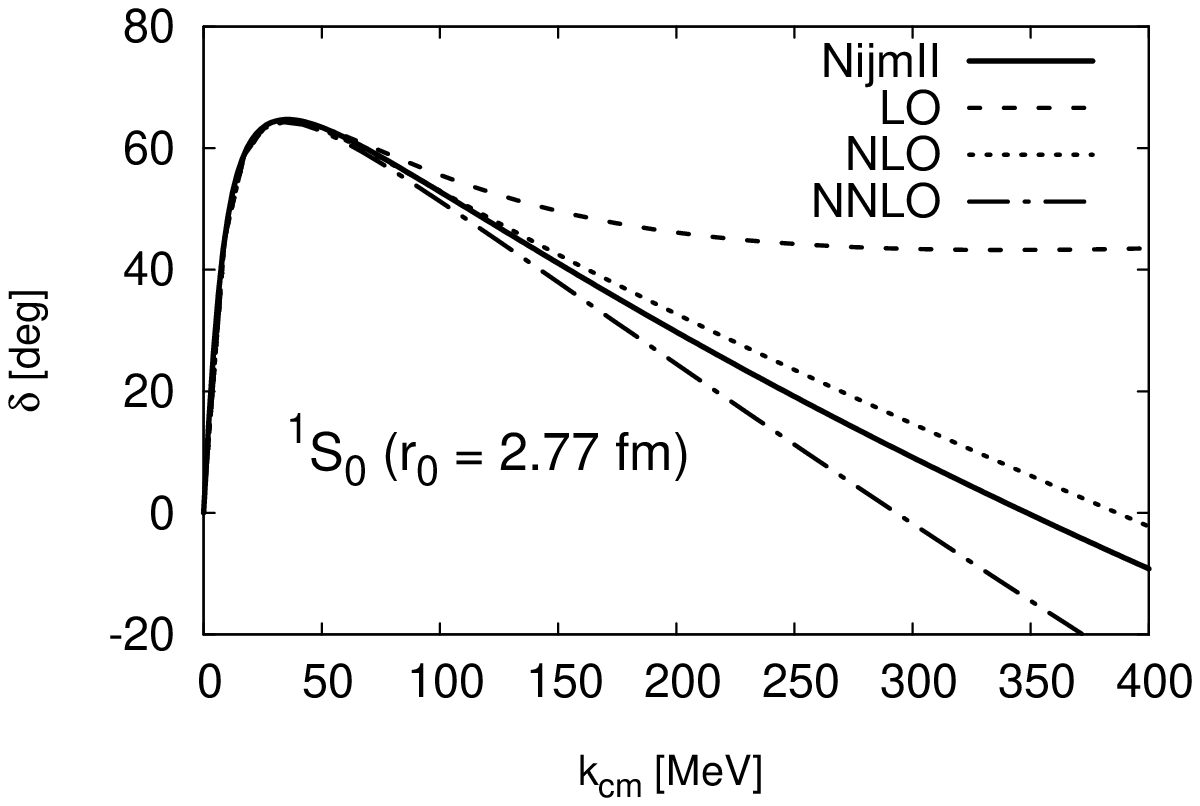}
\caption{Renormalized $^1S_0$ phase shifts (in degrees) for chiral
LO,NLO and NNLO potentials fixing $\alpha_0= -23.74 {\rm fm} $ (Left
panel) or $\alpha_0= -23.74 {\rm fm} $ and $r_0 = 2.77 {\rm fm}
$(Right panel) as input parameters.  The data are
from~\cite{Stoks:1994wp}.}
\label{fig:1C+2C}
\end{center}
\end{figure}

The generalization of the above method to the singular chiral
potentials~\cite{Kaiser:1997mw,Rentmeester:1999vw} has been implemented
in~\cite{OPE+TPE}
with promising results for One- and Two Pion Exchange (OPE and
TPE). We illustrate again the case of pn scattering in the
$^1S_0$-channel.  For the simplest situation with one RC,
corresponding to fix the scattering length as an independent
parameter, the method outlined above may be directly applied to
singular potentials {\it provided} they are attractive, i.e. $V(r) \to
-C_n/r^n$ with $n \ge 2 $~\footnote{If the potential was singular and
repulsive one cannot fix any low energy parameters; doing so yields
non-converging phase shifts.}.  The result for zero energy wave
functions as well as the effective range can be seen at
Fig.~\ref{fig:s-wave}. NNLO corresponds to the TPE potential of
Ref.~\cite{Kaiser:1997mw}. As we see the Nijmegen result $r_0=2.67
{\rm fm}$ is {\it almost} saturated by the TPE potential yielding $r_0
= 2.87 {\rm fm}$ already at $r_c \sim 0.5 {\rm fm}$. Calculations with
TPE to N3LO with one RC show convergence but no
improvement~\cite{Entem:2007jg} without or with $\Delta$ explicit
degrees of freedom. Thus, some physics is missing, perhaps 3$\pi$
effects. If, in addition to $\alpha_0$, we want to fix $r_0= 2.67
{\rm fm}$~\cite{Stoks:1994wp} as a RC we must solve Eqs.~(\ref{eq:u0})
and (\ref{eq:u2}). The matching condition at the boundary $r=r_c$
becomes energy dependent~\cite{PavonValderrama:2007nu}
\begin{eqnarray}
\frac{u'_p (r_c)}{u_p(r_c)} = \frac{u'_0 (r_c) + p^2 u'_2 (r_c)+ \dots}{
u_0(r_c)+ p^2 u_2(r_c)+ \dots} \, .
\label{eq:Lp}
\end{eqnarray} 
The generalization to arbitrary order is straightforward.  For $N$
RC's we have $u_p (r)= \sum_{n=0}^N p^{2n} u_{2n} (r)$ and using the
natural extension of the matching relation in Eq.~(\ref{eq:Lp}) as
well as the superposition principle of boundary conditions one can
show the following formula 
\begin{eqnarray} 
p \cot \delta_0 (p) = \frac{ \sum_{n=0}^N a_n {\cal A}_n (p, r_c)} {
\sum_{n=0}^N a_n {\cal B}_n (p, r_c)} \, , 
\label{eq:pcotd}
\end{eqnarray} 
where the coefficients $a_n$ can be related to the effective range
parameters $ a_0 = 1 $, $a_1 = - 1/\alpha_0 $, $a_2 = r_0 $,$ a_3 =
v_2 $ etc.  and ${\cal A}_n (p, r_c)$ and ${\cal B}_n (p, r_c)$ are
functions which are finite in the limit $r_c \to 0$ and depend {\it
solely } on the potential. In Eq.~(\ref{eq:pcotd}) the dependence on
the low energy parameters used as input is displayed explicitly and
can be completely separated from the long range
potential~\cite{PavonValderrama:2007nu}. The coupled channel case can
be analyzed in terms of eigenpotentials although the result is
cumbersome. In Fig.~\ref{fig:1C+2C} we
show the phase shitf for the $^1S_0$ channel when the potential is
considered at LO, NLO and NNLO and either one RC (fixing $\alpha_0$)
or two RC's (fixing $\alpha_0$ and $r_0$) are considered. LO+1C,
NLO+2C and NNLO+2C fix the same number or RC's as LO, NLO and NNLO of
the Weinberg counting respectively. As we see, our NNLO+2C does not
improve over NLO+2C.

It is worth mentioning that the innocent-looking energy dependent
matching condition, Eq.~(\ref{eq:Lp}), is quite unique since this is
the only representation guaranteeing finiteness of results for
singular potentials~\cite{PavonValderrama:2007nu}. Polynomial
expansions in $p^2$ such as suggested e.g. in
Ref.~\cite{Rentmeester:1999vw} do not work for $r_c\to 0$. A virtue of
the coordinate over momentum space is that these results can be
deduced analytically. For instance, the equivalent representation of
Eq.(\ref{eq:pcotd}) in momentum space  may likely exist, but is so far
unknown. Actually, the usual polynomial representation of short
distance interactions in momentum space $V_S(k',k) = C_0 + C_2 (k^2 +
k'^2 ) + \dots $ of standard NLO and NNLO Weinberg counting is
renormalizable only when $C_2 \to 0$ for $\Lambda \to
\infty$~\cite{Entem:2007jg}.

\section{Renormalization of the Deuteron}

In the $^3S_1-{}^3D_1$ channel, the relative proton-neutron state 
for negative energy is described by the coupled equations
\begin{eqnarray}
\begin{pmatrix}
- \frac{d^2}{dr^2} + M_N V_s (r) & M_N V_{sd} (r) \\
M_N V_{sd}(r) & - \frac{d^2}{dr^2} + \frac{6}{r^2} + M_N V_d (r) 
\end{pmatrix}
\begin{pmatrix}
u \\ w  
\end{pmatrix} 
= - \gamma^2 
\begin{pmatrix}
u \\ w  
\end{pmatrix} \, . 
\label{eq:sch_coupled} 
\end{eqnarray}
Here $ \gamma = \sqrt{M_M B} $, with $B=2.24 {\rm MeV}$ is the
deuteron binding energy and $u (r)$ and $w (r)$ are S- and D-wave
reduced wave functions respectively. At long distances they satisfy,
\begin{eqnarray}
\begin{pmatrix} 
u \\ 
w 
\end{pmatrix} \to  A_S\,e^{-\gamma r} \, 
\begin{pmatrix}
1 \\ 
\eta \left[1 + \frac{3}{\gamma r} + \frac{3}{(\gamma r)^2} \right] 
\end{pmatrix} \, .
\label{eq:bcinfty_coupled} 
\end{eqnarray}
where $\eta$ is the asymptotic D/S ratio parameter and $A_S$ is the
asymptotic normalization factor, which is such that the deuteron wave
functions are normalized to unity. The OPE $^3S_1-^3D_1$ potential is
given by $ M_N V_s = U_C$ , $ M_N V_{sd} = 2 \sqrt{2} U_T$,$ M_N V_d
= U_C - 2 U_T$ where for $r \ge r_c > 0$ we have
\begin{eqnarray}
U_C = -\frac{m_\pi^2 M_N g_A^2 }{16 \pi f_\pi^2 } \frac{e^{-m_\pi r 
}}{r} \, , \qquad U_T =&  U_C \left( 1 + \frac3{m_\pi r}+ \frac3{(m_\pi 
r)^2} \right) \, .  
\end{eqnarray} 
The tensor force generates a $1/r^3$ singularity at the origin in
coupled channel space. This behavior of the potential is strong
enough to overcome the centrifugal barrier at short distances, thus
modifying the usual threshold behavior of the wave functions. The
interesting aspect of this potential is that after diagonalization it
has one positive (repulsive) and one (negative) attractive
eigenvalue. The proper normalization of the wave functions in the
limit $r_c \to 0$ implies that one can only fix one free parameter,
e.g. the deuteron binding energy~\cite{OPE+TPE}.  Other properties may
be predicted, for instance one gets $\eta_{\rm OPE}=0.0263$
(exp. $0.0256(4)$).
The TPE chiral potentials of Ref.~\cite{Kaiser:1997mw} have also been
renormalized~\cite{OPE+TPE}, yielding a rather satisfactory picture of
the deuteron. The results described here have been reproduced in
momentum space~\cite{PVNRAP}. The required cut-off in momentum space
is larger than a naive estimate $r_c \sim 1/ \Lambda$ because the
regularization influences both the counterterms as well as the
potential. Deuteron form factors, probing some off-shellness of the
potential, have been computed describing surprisingly well the data up
to momenta $q \sim 800 {\rm MeV}$ when LO currents are
considered~\footnote{See talk of D. R. Phillips in this conference}.

\section{Power counting and renormalization}

The question on how a sensible hierarchy for NN interactions should be
organized remains so far open, because it is not obvious if one should
renormalize or not and
how~\cite{OPE+TPE,Nogga:2005hy,Epelbaum:2006pt,Birse:2005um}. However,
for a given long distance potential, we know {\it whether} and, in
positive case, {\it how} this can be made compatible with the desired
short distance insensitivity~\cite{OPE+TPE}. Not all chiral
interactions fit into this scheme, and thus it is sometimes preferred
to keep finite cut-offs despite results being often strongly dependent
on the choice at scales $r_c \sim 0.5 - 1 {\rm fm}$ similar to the
ones we want to probe in NN
scattering~\cite{OPE+TPE,Nogga:2005hy}. Renormalizability of chiral
potentials developing a singularity such as Eq.~(\ref{eq:pot-sing})
requires that one must choose the regular solution in which case the
wave function behaves as $u_p (r) \sim ( r 4 \pi f_\pi)^{\frac{2n+m}4} $ and
thus increasing insensitivity is guaranteed as the power of the
singularity increases. Converging renormalized TPE calculations show
insensitivity for reasonable scales of $r_c \sim 0.5 {\rm
fm}$~\cite{OPE+TPE}.

The Weinberg counting based in a heavy baryon approach at
LO~\cite{Epelbaum:2005pn} for $^1S_0$ and $^3S_1-^3D_1$ states turns
out to be renormalizable. There is at present no necessity argument
why this ought to be so, for the simple reason that power counting
does not anticipate the sign of the interaction at short
distances. When one goes to NLO the short distance $1/r^5$ singular
repulsive character of the potential makes the deuteron
unbound~\cite{OPE+TPE}. Finally, NNLO potentials diverge as $-1/r^6$
and are, again, compatible with Weinberg counting in the
deuteron\cite{OPE+TPE}. More failures have been reported in
Refs.~\cite{Nogga:2005hy,Entem:2007jg}. Relativistic potentials
subjected to different power counting have been renormalized in
Ref.~\cite{Higa:2007gz} yielding much less counterterms due to their
different short distance $1/r^7$ singularities and slightly better
overall description, although the $^1S_0$ phase is not improved as
compared to the heavy baryon formulation. These complications in the
more fundamental chiral potentials contrast with the simplicity of the
$\sigma+\pi$ OBE renormalized results (see Figs.~\ref{fig:sigma} and
\ref{fig:1C+2C}).

In the present state of affairs a clue might come from a remarkable
analogy between the NN interaction in the chiral quark model and the
Van der Waals molecular interactions in the Born-Oppenheimer
approximation~\cite{OPE+TPE}.  For non-relativistic constituent quarks
the direct NN interaction is provided by the convoluted OPE
quark-quark potential. Second order perturbation theory in OPE
among quarks generates TPE between nucleons yielding 
\begin{eqnarray}
V_{NN} =\langle NN | V_{\rm OPE} | NN \rangle + \sum_{HH' \neq NN
}\frac{ |\langle NN | V_{\rm OPE} | HH' \rangle |^2}{ E_{NN}-E_{HH'} }+
\dots 
\end{eqnarray} 
When $H H'=N \Delta$ and $H H'=\Delta \Delta$ this resembles
Ref.~\cite{Kaiser:1998wa} which for $ 2 {\rm fm} < r < 3 {\rm fm }$
behaves as $\sigma$ exchange with $m_\sigma=550 {\rm MeV}$ and
$g_{\sigma NN} = 9.4$.  Moreover, the second order perturbative
character suggests that the potential becomes singular $\sim 1 / r^6 $
and attractive, necessarily being renormalizable with an arbitrary
number of counterterms through energy dependent boundary
conditions~\cite{PavonValderrama:2007nu}. Clearly, the renormalization
of such a scheme where the $N\Delta$ splitting is treated as a small
scale deserves further investigation~\cite{PVRA07}.

\section{Conclusion}

Renormalization is the mathematical implementation of the appealing
physical requirement of short distance insensitivity and hence a
convenient tool to search for model independent results.  In a
non-perturbative setup such as the NN problem, renormalization imposes
rather tight constraints on the interplay between the unknown short
distance physics and the perturbatively computable long distance
interactions.  This viewpoint provides useful insights and it is
within such a framework that we envisage a systematic and model
independent description of the NN force based on chiral interactions.

\section*{Acknowledgments}

We thank R. Higa, D. Entem, R. Machleidt, A. Nogga and D. R. Phillips
for collaboration and the Spanish DGI and FEDER funds grant
no. FIS2005-00810, Junta de Andaluc{\'\i}a grants no.FQM225-05, EU
Integrated Infrastructure Initiative Hadron Physics Project grant 
no. RII3-CT-2004-506078, DFG (SFB/TR 16) and Helmholtz Association
grant no. VH-NG-222 for support.


\end{document}